\documentstyle[twoside]{article}

\catcode`\@=11
\long\def\@makefntext#1{
\protect\noindent \hbox to 3.2pt {\hskip-.9pt 
$^{{\eightrm\@thefnmark}}$\hfil}#1\hfill}		

\def\thefootnote{\fnsymbol{footnote}}
\def\@makefnmark{\hbox to 0pt{$^{\@thefnmark}$\hss}}	
	
\def\ps@myheadings{\let\@mkboth\@gobbletwo
\def\@oddhead{\hbox{}
\rightmark\hfil\eightrm\thepage}
\def\@oddfoot{}\def\@evenhead{\eightrm\thepage\hfil
\leftmark\hbox{}}\def\@evenfoot{}
\def\sectionmark##1{}\def\subsectionmark##1{}}

\oddsidemargin=\evensidemargin
\renewcommand{\thefootnote}{\fnsymbol{footnote}}

\newcounter{sectionc}\newcounter{subsectionc}\newcounter{subsubsectionc}
\renewcommand{\section}[1] {\vspace{12pt}\addtocounter{sectionc}{1}
\setcounter{subsectionc}{0}\setcounter{subsubsectionc}{0}\noindent
	{\tenbf\thesectionc. #1}\par\vspace{5pt}}
\renewcommand{\subsection}[1] {\vspace{12pt}\addtocounter{subsectionc}{1}
	\setcounter{subsubsectionc}{0}\noindent
	{\bf\thesectionc.\thesubsectionc. {\kern1pt \bfit #1}}\par\vspace{5pt}}
\renewcommand{\subsubsection}[1] {\vspace{12pt}\addtocounter{subsubsectionc}{1}
	\noindent{\tenrm\thesectionc.\thesubsectionc.\thesubsubsectionc.
	{\kern1pt \tenit #1}}\par\vspace{5pt}}
\newcommand{\nonumsection}[1] {\vspace{12pt}\noindent{\tenbf #1}
	\par\vspace{5pt}}

\newcounter{appendixc}
\newcounter{subappendixc}[appendixc]
\newcounter{subsubappendixc}[subappendixc]
\renewcommand{\thesubappendixc}{\Alph{appendixc}.\arabic{subappendixc}}
\renewcommand{\thesubsubappendixc}
	{\Alph{appendixc}.\arabic{subappendixc}.\arabic{subsubappendixc}}

\renewcommand{\appendix}[1] {\vspace{12pt}
        \refstepcounter{appendixc}
        \setcounter{figure}{0}
        \setcounter{table}{0}
        \setcounter{lemma}{0}
        \setcounter{theorem}{0}
        \setcounter{corollary}{0}
        \setcounter{definition}{0}
        \setcounter{equation}{0}
        \renewcommand{\thefigure}{\Alph{appendixc}.\arabic{figure}}
        \renewcommand{\thetable}{\Alph{appendixc}.\arabic{table}}
        \renewcommand{\theappendixc}{\Alph{appendixc}}
        \renewcommand{\thelemma}{\Alph{appendixc}.\arabic{lemma}}
        \renewcommand{\thetheorem}{\Alph{appendixc}.\arabic{theorem}}
        \renewcommand{\thedefinition}{\Alph{appendixc}.\arabic{definition}}
        \renewcommand{\thecorollary}{\Alph{appendixc}.\arabic{corollary}}
        \renewcommand{\theequation}{\Alph{appendixc}.\arabic{equation}}
        \noindent{\tenbf Appendix \theappendixc #1}\par\vspace{5pt}}
\newcommand{\subappendix}[1] {\vspace{12pt}
        \refstepcounter{subappendixc}
        \noindent{\bf Appendix \thesubappendixc. {\kern1pt \bfit #1}}
	\par\vspace{5pt}}
\newcommand{\subsubappendix}[1] {\vspace{12pt}
        \refstepcounter{subsubappendixc}
        \noindent{\rm Appendix \thesubsubappendixc. {\kern1pt \tenit #1}}
	\par\vspace{5pt}}

\newcommand{\textlineskip}{\baselineskip=13pt}
\newcommand{\smalllineskip}{\baselineskip=10pt}

\def\eightcirc{
\begin{picture}(0,0)
\put(4.4,1.8){\circle{6.5}}
\end{picture}}
\def\eightcopyright{\eightcirc\kern2.7pt\hbox{\eightrm c}}

\newcommand{\copyrightheading}[1]
	{\vspace*{-2.5cm}\smalllineskip{\flushleft
	{\footnotesize International Journal of Modern Physics A, #1}\\
	{\footnotesize $\eightcopyright$\, World Scientific Publishing
	 Company}\\
	 }}

\def\abstracts#1#2#3{{
	\centering{\begin{minipage}{4.5in}\baselineskip=10pt\footnotesize
	\parindent=0pt #1\par
	\parindent=15pt #2\par
	\parindent=15pt #3
	\end{minipage}}\par}}

\renewenvironment{thebibliography}[1]
	{\frenchspacing
	 \ninerm\baselineskip=11pt
	 \begin{list}{\arabic{enumi}.}
	{\usecounter{enumi}\setlength{\parsep}{0pt}
	 \setlength{\leftmargin 12.7pt}{\rightmargin 0pt} 
	 \setlength{\itemsep}{0pt} \settowidth
	{\labelwidth}{#1.}\sloppy}}{\end{list}}

\newcounter{itemlistc}
\newcounter{romanlistc}
\newcounter{alphlistc}
\newcounter{arabiclistc}

\newcommand{\fcaption}[1]{
        \refstepcounter{figure}
        \setbox\@tempboxa = \hbox{\footnotesize Fig.~\thefigure. #1}
        \ifdim \wd\@tempboxa > 5in
           {\begin{center}
        \parbox{5in}{\footnotesize\smalllineskip Fig.~\thefigure. #1}
            \end{center}}
        \else
             {\begin{center}
             {\footnotesize Fig.~\thefigure. #1}
              \end{center}}
        \fi}

\newcommand{\tcaption}[1]{
        \refstepcounter{table}
        \setbox\@tempboxa = \hbox{\footnotesize Table~\thetable. #1}
        \ifdim \wd\@tempboxa > 5in
           {\begin{center}
        \parbox{5in}{\footnotesize\smalllineskip Table~\thetable. #1}
            \end{center}}
        \else
             {\begin{center}
             {\footnotesize Table~\thetable. #1}
              \end{center}}
        \fi}

\def\@citex[#1]#2{\if@filesw\immediate\write\@auxout
	{\string\citation{#2}}\fi
\def\@citea{}\@cite{\@for\@citeb:=#2\do
	{\@citea\def\@citea{,}\@ifundefined
	{b@\@citeb}{{\bf ?}\@warning
	{Citation `\@citeb' on page \thepage \space undefined}}
	{\csname b@\@citeb\endcsname}}}{#1}}

\newif\if@cghi
\def\cite{\@cghitrue\@ifnextchar [{\@tempswatrue
	\@citex}{\@tempswafalse\@citex[]}}
\def\citelow{\@cghifalse\@ifnextchar [{\@tempswatrue
	\@citex}{\@tempswafalse\@citex[]}}
\def\@cite#1#2{{$\null^{#1}$\if@tempswa\typeout
	{IJCGA warning: optional citation argument
	ignored: `#2'} \fi}}

\def\pmb#1{\setbox0=\hbox{#1}
	\kern-.025em\copy0\kern-\wd0
	\kern.05em\copy0\kern-\wd0
	\kern-.025em\raise.0433em\box0}

\def\fnt#1#2{\footnotetext{\kern-.3em
	{$^{\mbox{\scriptsize #1}}$}{#2}}}

\def\fpage#1{\begingroup
\voffset=.3in
\thispagestyle{empty}\begin{table}[b]\centerline{\footnotesize #1}
	\end{table}\endgroup}

\def\runninghead#1#2{\pagestyle{myheadings}
\markboth{{\protect\footnotesize\it{\quad #1}}\hfill}
{\hfill{\protect\footnotesize\it{#2\quad}}}}
\font\tenrm=cmr10
\font\tenit=cmti10
\font\tenbf=cmbx10
\font\bfit=cmbxti10 at 10pt
\font\ninerm=cmr9

\font\eightrm=cmr8

\def\qed{\hbox{${\vcenter{\vbox{			
   \hrule height 0.4pt\hbox{\vrule width 0.4pt height 6pt
   \kern5pt\vrule width 0.4pt}\hrule height 0.4pt}}}$}}

\renewcommand{\thefootnote}{\fnsymbol{footnote}}	

\newcount\@tempcntc
\def\@citex[#1]#2{\if@filesw\immediate\write\@auxout{\string\citation{#2}}\fi
  \@tempcnta\z@\@tempcntb\m@ne\def\@citea{}\@cite{\@for\@citeb:=#2\do
    {\@ifundefined
       {b@\@citeb}{\@citeo\@tempcntb\m@ne\@citea\def\@citea{,}{\bf ?}\@warning
       {Citation `\@citeb' on page \thepage \space undefined}}%
    {\setbox\z@\hbox{\global\@tempcntc0\csname b@\@citeb\endcsname\relax}%
     \ifnum\@tempcntc=\z@ \@citeo\@tempcntb\m@ne
       \@citea\def\@citea{,}\hbox{\csname b@\@citeb\endcsname}%
     \else
      \advance\@tempcntb\@ne
      \ifnum\@tempcntb=\@tempcntc
      \else\advance\@tempcntb\m@ne\@citeo
      \@tempcnta\@tempcntc\@tempcntb\@tempcntc\fi\fi}}\@citeo}{#1}}
\def\@citeo{\ifnum\@tempcnta>\@tempcntb\else\@citea\def\@citea{,}%
  \ifnum\@tempcnta=\@tempcntb\the\@tempcnta\else
   {\advance\@tempcnta\@ne\ifnum\@tempcnta=\@tempcntb \else \def\@citea{-}\fi
    \advance\@tempcnta\m@ne\the\@tempcnta\@citea\the\@tempcntb}\fi\fi}

\def\etal{{\it et al.}}
\def\PL{Phys. Lett.}
\def\PRL{Phys. Rev. Lett.}

\newcommand{\eqn}[1]{(\ref{#1})}
\newcommand{\be}{\begin{equation}}
\newcommand{\ee}{\end{equation}}

\newcommand{\ba}{\begin{array}{c}}
\newcommand{\bat}{\begin{array}{cc}}
\newcommand{\ea}{\end{array}}
\newcommand{\beqn}{\begin{eqnarray}}
\newcommand{\eeqn}{\end{eqnarray}}

\newcommand{\bi}{\begin{itemize}}
\newcommand{\ei}{\end{itemize}}

\def\eps{\varepsilon}

\newcommand{\cA}{{\cal A}}

\begin{document}

\runninghead{The role of final state interactions in $\eps'/\eps$}{
The role of final state interactions in $\eps'/\eps$}

\normalsize\textlineskip
\thispagestyle{empty}
\setcounter{page}{1}

\copyrightheading{}			

\vspace*{0.88truein}

\fpage{1}
\centerline{\bf THE ROLE OF FINAL STATE INTERACTIONS IN
      $\varepsilon'/\varepsilon$
   %
 }
\vspace*{0.37truein}
\centerline{\footnotesize ELISABETTA PALLANTE}
\vspace*{0.015truein}
\centerline{\footnotesize\it SISSA, Via Beirut 2-4, I-34013 Trieste, Italy}
\vspace*{10pt}
\centerline{\footnotesize ANTONIO PICH, \  IGNAZIO SCIMEMI}
\vspace*{0.015truein}
\centerline{\footnotesize\it
         IFIC, Universitat de Val\`encia -- CSIC,  
  Apt. Correus 22085, E--46071 Val\`encia, Spain}
\vspace*{0.225truein}

\vspace*{0.21truein}
\abstracts{
The Standard Model prediction for $\varepsilon'/\varepsilon$ is updated,
taking into account the chiral loop corrections induced by final state
interactions. The resulting value,
$\varepsilon'/\varepsilon = (17\pm 6)\times 10^{-4}$,
is in good agreement with present measurements.}{}{}

\textheight=7.8truein
\setcounter{footnote}{0}
\renewcommand{\thefootnote}{\alph{footnote}}

\vspace*{1pt}\textlineskip	
\section{Introduction}	\label{sec:introduction}
\vspace*{-0.5pt}
\noindent
The CP--violating ratio  $\varepsilon'/\varepsilon$  constitutes
a fundamental test for our understanding of flavour--changing
phenomena.
The present experimental world average,\cite{ktev:99}
${\rm Re} \left(\varepsilon'/\varepsilon\right) =
(19.3 \pm 2.4) \cdot 10^{-4}$,
provides clear evidence for a non-zero value and,
therefore, the existence of direct CP violation.

The theoretical prediction has been rather controversial since
different groups, using different models or approximations,
have obtained different
results.\cite{PP:00a,PP:00b,PPS:00,munich,rome,trieste,dortmund,BP:00}
In terms of the $K\to\pi\pi$ isospin amplitudes,
$\cA_I = A_I \, e^{i\delta_I}$ ($I=0,2$),
\be
{\varepsilon^\prime\over\varepsilon} =
\; e^{i\Phi}\; {\omega\over \sqrt{2}\vert\eps\vert}\;\left[
{\mbox{Im}A_2\over\mbox{Re} A_2} - {\mbox{Im}A_0\over \mbox{Re} A_0}
 \right] \, ,
\qquad\qquad
\Phi \approx \delta_2-\delta_0+\frac{\pi}{4}\approx 0 \, ,
\ee
where
$\omega = \mbox{Re} A_2/\mbox{Re} A_0 \approx 1/22$.
The CP--conserving amplitudes $\mbox{Re} A_I$, their ratio
$\omega$ and $\eps$ are usually set to their experimentally
determined values. A theoretical calculation is then only needed
for the quantities $\mbox{Im} A_I$.

Since $M_W\gg M_K$, there are large short--distance logarithmic
contributions which can be summed up using the Operator Product
Expansion and the renormalization group.\cite{buras1,ciuc1}
To predict the physical amplitudes one also needs to
compute long--distance hadronic matrix elements of light
four--quark operators $Q_i$. They are usually parameterized
in terms of the so-called bag parameters $B_i$, which measure them
in units of their vacuum insertion approximation values.

To a very good approximation, the Standard Model prediction for
$\varepsilon'/\varepsilon$ can be written (up to global factors)
as\cite{munich}
\be
{\varepsilon'\over\varepsilon} \sim
\left [ B_6^{(1/2)}(1-\Omega_{IB}) - 0.4 \, B_8^{(3/2)}
 \right ]\, , \qquad\quad
\Omega_{IB} = {1\over \omega}
{(\mbox{Im}A_2)_{IB}
\over \mbox{Im}A_0} \, .
\label{EPSNUM}
\ee
Thus, only two operators are numerically relevant:
the QCD penguin operator $Q_6$ governs $\mbox{Im}A_0$
($\Delta I=1/2$), while $\mbox{Im}A_2$ ($\Delta I=3/2$)
is dominated by the electroweak penguin operator $Q_8$.
The parameter $\Omega_{IB}$
takes into account isospin breaking corrections;
the value $\Omega_{IB}=0.25$
was usually adopted in all calculations.\cite{Omega}
Together with $B_i\sim 1$, this produces a numerical cancellation
leading to values of $\eps'/\eps\sim 7\times 10^{-4}$.
This number has been slightly increased by a
recent Chiral Perturbation Theory ($\chi$PT) calculation at $O(p^4)$
which finds $\Omega_{IB}= 0.16\pm 0.03$.\cite{EMNP:00}

\section{Chiral Loop Corrections}
\label{sec:ChPT}
\noindent
Chiral symmetry determines the low--energy hadronic realization of the
operators $Q_i$,
through a perturbative expansion in powers of momenta and quark masses.
The corresponding chiral couplings can be calculated in the
large--$N_C$ limit of QCD. The usual input values
$B_8^{(3/2)}\approx B_6^{(1/2)}=1$ correspond
to the lowest--order approximation in both the $1/N_C$ and
$\chi$PT expansions.

The lowest--order calculation does not provide any strong phases
$\delta_I$. Those phases originate in the
final rescattering of the two pions and, therefore, are generated by
higher--order chiral loops.
Analyticity and unitarity require the presence of a corresponding
dispersive effect in the moduli of the isospin amplitudes.
Since the S--wave strong phases are quite large,
specially in the isospin--zero case,
one should expect large unitarity corrections.

The one--loop analyses of $K\to 2 \pi$ show in fact that pion
loop diagrams provide an important enhancement of the $\cA_0$
amplitude.\cite{KA91}
This chiral loop correction destroys the accidental numerical
cancellation in eq.~\eqn{EPSNUM}, generating a sizeable enhancement
of the $\eps'/\eps$ prediction.\cite{PP:00a}
The large one--loop correction to $\cA_0$ has its origin in the
strong final state interaction (FSI) of the two pions in S--wave,
which generates large infrared logarithms involving the light
pion mass.\cite{PP:00b}
Using analyticity and unitarity constraints,
these logarithms can be exponentiated to all orders in
the chiral expansion.\cite{PP:00a,PP:00b}
For the CP--conserving amplitudes, the result can be written as
\be\label{eq:OMNES_WA}
\cA_I \, =\,  \left(M_K^2-M_\pi^2\right) \; a_I(M_K^2) \, =\,
\left(M_K^2-M_\pi^2\right) \; \Omega_I(M_K^2,s_0) \; a_I(s_0)\, ,
\ee
where $a_I(s)$ denote reduced off-shell amplitudes with
$s\equiv \left(p_{\pi_1}+p_{\pi_2}\right)^2$ and
\be\label{eq:omega}
\Omega_I(s,s_0) \,\equiv\, e^{i\delta_I(s)}\; \Re_I(s,s_0) \, =\,
 \exp{\left\{ {(s-s_0)\over\pi}\int
{dz\over (z-s_0)} {\delta_I(z)\over (z-s-i\epsilon)}\right\}}
\ee
provides an evolution of $a_I(s)$ from an arbitrary
low--energy point $s_0$ to $s=M_K^2$.
The physical amplitude $a_I(M_K^2)$ is of course independent of $s_0$.

Taking the chiral prediction for $\delta_I(z)$ and expanding
the exponential to first order,
one just reproduces the one--loop $\chi$PT result.
Eq.~\eqn{eq:omega} allows us to get a much more accurate
prediction, by taking $s_0$ low enough that the $\chi$PT corrections
to $a_I(s_0)$ are
small and exponentiating the large logarithms with the
Omn\`es factor $\Omega_I(M_K^2,s_0)$.
Moreover, using the experimental phase-shifts in the dispersive
integral one achieves an all--order resummation of FSI effects.
The numerical accuracy of this exponentiation has been successfully
tested through an analysis of the scalar pion form factor,\cite{PP:00b}
which has identical FSI than $\cA_0$.

\section{Numerical Predictions}
\label{sec:numerics}
\noindent
At $s_0 =0$, the chiral corrections are rather small.
To a very good approximation,\cite{PPS:00} we can just multiply
the tree--level $\chi$PT result for $a_I(0)$
with the experimentally determined Omn\`es exponentials:\cite{PP:00b}
\be
\Re_0\equiv\Re_0(M_K^2,0) =  1.55 \pm 0.10\, ,
\qquad\qquad
\Re_2\equiv\Re_2(M_K^2,0) =  0.92 \pm 0.03\, .
\ee
Thus, \
$B_6^{(1/2)} \approx \Re_0\times
\left. B_6^{(1/2)}\right|_{N_C\to\infty} = 1.55$, \
$B_8^{(3/2)} \approx\Re_2\times
\left. B_8^{(3/2)}\right|_{N_C\to\infty} \approx  0.92$\
and\
$\Omega_{IB} \approx 0.16 \times\Re_2/\Re_0= 0.09$.
This agrees with the result \ $\Omega_{IB} = 0.08\pm 0.05$,
obtained recently with an explicit chiral loop calculation.\cite{MW:00}

The large FSI correction to the $I=0$ amplitude gets reinforced
by the mild suppression of the $I=2$ contributions. The net effect
is a large enhancement of $\eps'/\eps$ by a factor 2.4,
pushing the predicted central value from\cite{munich,rome}
$7\times 10^{-4}$ to\cite{PP:00b} $17\times 10^{-4}$.
A more careful analysis, taking into account all hadronic and
quark--mixing inputs gives the Standard Model prediction:\cite{PPS:00}
\be\label{eq:SMpred}
\varepsilon'/\varepsilon = (17\pm 6) \times 10^{-4}\, ,
\ee
which compares well with the present experimental world average.

\nonumsection{Acknowledgements}
\noindent
This work has been supported by the ECC, TMR Network
$EURODAPHNE$ (ERBFMX-CT98-0169), and by
DGESIC (Spain) under grant No. PB97-1261.

\nonumsection{References}
 

\end{document}